\documentclass[aps,twocolumn,amsmath,amssymb]{revtex4} 
\usepackage[english]{babel}
\usepackage{latexsym}
\usepackage{graphics}
\usepackage{subfigure}
\usepackage{epsfig}

\begin{document}


\title{Dynamical phase transitions  and temperature induced quantum correlations in an infinite spin chain}


\author{Aditi Sen(De)\(^{(1),(2)}\), 
Ujjwal Sen\(^{(1),(2)}\), and Maciej Lewenstein\(^{(1),(2),}\)\footnote{Also at Instituci\'o Catalana de Recerca i Estudis Avan{\c c}ats.}
}

\affiliation{\(^{(1)}\)ICFO-Institut de Ci\`encies Fot\`oniques, Jordi Girona 29, Edifici Nexus II, E-08034 Barcelona, Spain\\
\(^{(2)}\)Institut f\"ur Theoretische Physik, 
Universit\"at Hannover, D-30167 Hannover,
Germany}

\begin{abstract}

We study the dynamics of entanglement in the infinite asymmetric \(XY\) spin chain, in an applied transverse field.
%
The system is prepared in a thermal equilibrium state (ground state at zero temperature) at the initial 
instant, and it starts evolving after the transverse field is completely turned off.  
We investigate the evolved state of the chain at a given fixed time, 
and  show that the nearest neighbor entanglement in the chain
exhibits a critical behavior (which we call a dynamical phase transition),
controlled by the initial value of the transverse field.
The character of the dynamical phase transition is qualitatively different 
for short and long evolution times. We  also find a \emph{nonmonotonic} behavior of entanglement with respect to the temperature 
of the initial equilibrium state. Interestingly, the region of the initial field for which we obtain a 
nonmonotonicity of entanglement (with respect to temperature)
is directly related to the position and character 
of the dynamical phase transition in the model. 

\end{abstract}

\maketitle

\def\tr{{\rm tr}}
\def\com#1{{\tt [\hskip.5cm #1 \hskip.5cm ]}}

\section{Introduction}
\label{intro}

  Exploring the properties of entanglement 
has recently attracted a lot of interest, due to 
the usefulness  of entanglement in quantum information processing tasks \cite{NC}. Recently, 
 several authors have begun to study 
the properties of entanglement in real physical many body systems, 
such as cold atomic gases in optical lattices (e.g. \cite{optical_lattice}) or in trapped gaseous
 Bose-Einstein condensates (e.g. \cite{BEC}). It turns out that such investigations 
  help us to understand the  physics of quantum phase transitions
(e.g. \cite{Osterloh, Nielsen, salikh-qwerty, salikh-qwerty1, ent_length}). 
Moreover, they are important for
implementations of
 quantum computation, or other 
  quantum information processing tasks in such physical systems.

  Among the potential candidates
  for implementing quantum computation are various models of spin 
  systems that can be realised with ultracold atoms in optical lattices (see e.g. \cite{toolbox, proposal_korechhey}). 
  There is therefore a strong motivation behind the study of entanglement in spin systems 
(see e.g. \cite{Osterloh, Nielsen, salikh-qwerty, salikh-qwerty1, ent_length, Wootters, Indranidi, numerical, amader, Dagmar} and references 
therein). 
  Moreover, studies of entanglement in spin models help us to relate entanglement to the 
  fundamental concepts, such as quantum phase transitions.
  In particular, it
  was shown that near a phase transition in the ground state of an exactly solvable spin model in one
  dimension (Ising model in a transverse field), two-particle entanglement remains short ranged, 
while two-particle correlation length diverges
  \cite{Osterloh, Nielsen}. The behavior of bipartite, as well as multipartite 
  entanglement in the ground states  and (thermal) equilibrium states of spin rings and chains
  has been studied from several perspectives \cite{Osterloh, Nielsen, salikh-qwerty, Wootters, amader, Dagmar}. 
It has been shown that, using the concept of localizable entanglement \cite{ent_length, Ciracnew} (cf. \cite{hmmmm, amrao-achhi-re-bhai}),
bounded from above by  entanglement of assistance \cite{Bennett-assistance}, and from below by correlation functions,
one can define an entanglement correlation length that diverges at the criticality, 
and majorizes the standard correlation length (see also \cite{kachpoka}).

  Studies of entanglement of the time-evolved state of spin models has also been carried out
  \cite{Briegel_orey_orey,Sougato, talk_dilo_Osterloh, numerical, amader, meti-pukur}. In particular,
  implementation of the ``one-way quantum computer'' and short range teleportation \cite{BBCJPW} of an unknown state 
  has been proposed by using the dynamics of spin systems in 
  \cite{Briegel_orey_orey, Briegel_ebong, meti-pukur, Sougato, Sougato_ebong}. 
  In Ref. \cite{amader}, it was shown
  that the nearest-neighbor entanglement of the time-evolved state in an infinite 
spin chain (asymmetric \(XY\) model in a transverse field), after an initial disturbance,   
  does not approach its equilibrium value (nonergodicity of entanglement).  
   Previous studies of quantum dynamics of spin models after a rapid change of the field include Refs. \cite{Stephenexpt, Stephen}, while 
effects of a sudden switching of the interaction in arrays of oscillators were studied in Ref. \cite{Martin}.

In this paper, we study the 
dynamics of nearest-neighbor entanglement in the evolution of 
an infinite spin chain described by the asymmetric \(XY\) model in a transverse field (see Eq. (\ref{asolH}) below). 
We take the  initial state of the evolution to be the equilibrium state 
at zero temperature, and suddenly turn off the transverse field at zero time.  
The system is thus given an initial disturbance, and its properties are then studied 
at later times. 
We find that the nearest-neighbor entanglement in the evolved state at  a fixed time shows a 
criticality (which we call a dynamical phase transition (DPT)) 
 with respect to the transverse external field. 
We refer to the  region of the  initial transverse field for 
  which the entanglement is nonvanishing (vanishing),  at a fixed time, as the 
  ``entangled phase'' (``separable phase'').  
Interestingly, the 
nature of the DPT  depends on whether we 
are near, or far from the time of initial disturbance. Moreover, 
for values of the initial transverse field near the criticalities, as well as in the separable 
phase, and for short times, the nearest-neighbor entanglement 
shows \emph{nonmonotonicity} with respect to temperature. 
Accordingly, we call the criticalities 
as ``critical regions'', signifying that ``critical" effects persist for a small region around the critical value of
the transverse field.
Such nonmonotonicity of entanglement with respect to temperature was also 
found in Ref. \cite{Scheel}, in the Jaynes-Cummings model.
 Study of nonmonotonicity of entanglement with respect to temperature is 
interesting, as preservation of entanglement in a hostile environment is one of the 
main challenges in quantum computation and quantum information in general.
A common belief is that temperature is a form of noise, i.e. it destroys the subtle quantum correlations. 
However, we show that for 
the infinite spin chain modelled by the \(XY\) Hamiltonian, at least 
in some situations, entanglement is not always monotonically decreasing with temperature, and 
interestingly, this is related to the appearance of a dynamical phase transition in that model.

As we have mentioned above, we characterise the DPT in the asymmetric \(XY\) chain by magnetic field and temperature, the usual 
control parameters used in 
statistical physics for characterising phase transitions. However, note that we also use time (\(t\)) as one of our 
control parameters in the characterisation.
As we will see in this paper, significant change of behavior is observed in the DPTs, as we change the parameter \(t\).

 The paper is organised as follows. In Sec. \ref{XYmodel}, we will recollect some facts about
  the  \(XY\) spin 
 model and fix some notations. In order to explore the properties of entanglement, one has to fix the measure with which one 
 quantifies 
 entanglement. In Sec. \ref{entanglement}, we define the entanglement measure
 that will be considered in this paper. The concept of dynamical phase 
 transition is discussed in  Sec. \ref{sotti}. 
 The nonmonotonicity of entanglement of the evolved state
 with respect to temperature of the initial state, and its connection to the
 dynamical phase transition, are discussed in Sec. \ref{sec_asol}. 
 We summarize our results in the final section (Sec. \ref{discussion}).

\section{The \(XY\) model in the transverse field}
\label{XYmodel}

\subsection{Description of the model}

For a one-dimensional spin chain of spin 1/2 particles, 
a simple form of the  Hamiltonian with  nearest neighbor interactions is given by
\begin{equation}
\label{inthamil}
H_{int} = \sum_{i} ({\cal A} S_i^x S_{i+1}^x + {\cal B} S_i^y S_{i+1}^y + 
{\cal C} S_i^z S_{i+1}^z),
\end{equation}
where \(A\), \(B\), and \(C\) are coupling constants, and  \(S_i^x\), \(S_i^y\), \(S_i^z\) are spin 1/2 operators (one-half of the
Pauli matrices) at the \(i\)-th site. One  can also introduce an external
magnetic field in the Hamiltonian, so that 
the total Hamiltonian is 
\[
H = H_{int} - h(t) H_{mag},
\]
where \(h(t)\) is a time-dependent function, to be specified below.
To obtain  a non-trivial effect on the dynamics due to the field part of the Hamiltonian,
one must choose the magnetic field and other parameters in such a way that the interaction part and 
the field part of the total Hamiltonian 
do not commute.

A simple way to attain that is to choose the field part as 
\[H_{mag} = \sum_i S_i^z,\] and  \(A =  1 + \gamma\), \(B =  1 - \gamma\), \(C= 0\) (with
\(\gamma \neq 0\)).  
Therefore the total Hamiltonian that we study in this paper takes the form (\(\gamma \ne 0\))
\begin{equation}
\label{asolH}
H = \sum_{i} \left((1 + \gamma) S_i^x S_{i+1}^x + (1 - \gamma) S_i^y S_{i+1}^y\right) -  
h(t) \sum_i S_i^z. 
\end{equation}
This  Hamiltonian is called the asymmetric \(XY\) model in a transverse field.

Such a system can be realized in atomic gas in an optical lattice (e.g. \cite{toolbox, proposal_korechhey}). 
Note that the condition of nonvanishing anisotropy \(\gamma\) is required, 
as otherwise the field part commutes with the 
interaction part.
This model is exactly solvable by succesive Jordan-Wigner,
Fourier, and Bogoliubov transformations \cite{LSM}. 
We still have to specify the time dependence of the magnetic field, which we choose to be a step function.
Precisely, we choose \(h(t)\) to be 
%
\begin{equation}
h(t) = \Big\{\begin{array}{c} a,  \quad t \leq 0 \\ 0, \quad t  >  0  \end{array},
\nonumber
\end{equation} 
 with \(a \ne 0\).
The system is thus given an initial disturbance, as the field is turned off. The 
properties of the evolved state are then studied at later times.

We will mainly be interested in studying the dynamics of nearest neighbor entanglement  of the
evolved state of such model. The  state that we consider, evolves 
according to the Hamiltonian \(H\), given in Eq. (\ref{asolH}). But it also depends on the
initial state, from which it starts evolving. 
Let us denote the (thermal) equilibrium state at the initial time, and at absolute 
temperature \(T\) as \(\rho_\beta^{eq}\):
\[\rho_\beta^{eq} = \mbox{exp}[- \beta H(0)]/Z_\beta. \]
Here \(Z_\beta\) is the partition function, given by 
\[Z_\beta = \tr(\mbox{exp}[-\beta H(0)]), \]   
and \(\beta = \frac{1}{kT}\), where \(k\) is the Boltzmann constant. 
Henceforth, we set \(k =1\). 
In all cases studied here, we will be choosing an equilibrium state \(\rho^{eq}_\beta\)
as our initial state. In particular, we will consider the case of zero temperature, i.e. when
\(\beta \rightarrow \infty\).

\subsection{Single and two particle reduced density matrices}

Although our main intention is to study the behavior of nearest neighbor 
entanglement, we will also calculate the single-site property of magnetization in this model.
As we will see, the behavior of magnetization (in particular, its nonmonotonicity)
does not depend on whether we are near, or far from  the DPT discussed in this paper.
Let us therefore find out the single-site and two-site reduced
density matrices of the evolved state. Let us suppose that the evolution starts off from the 
initial state \(\rho^{eq}_\beta\) of the infinite chain, and let the evolved state of the 
infinite spin chain be denoted 
by \(\rho_\beta(t)\). Due to symmetry, all the single-site density matrices of the 
evolved state (at a particular instant \(t\), and for a particular temperature \(T\)) are equal. 
The same is true for the nearest neighbor density matrices (as for the other two-site density matrices).
 We denote them as 
\(\rho^1_\beta(t)\) and \(\rho^{12}_\beta(t)\), respectively.

Our system is an infinite spin chain of 
spin-1/2 particles, and so
  our single-site density matrix \(\rho^1_\beta(t)\) acts on the two-dimensional complex Hilbert space. 
  A general
  single qubit (two-dimensional quantum system) density matrix can be written as
  \[
  \rho^1_\beta (t)= \frac{1}{2} I + 2 M^z_\beta(t) S^z + 2 M^x_\beta(t) S^x + 2 M^y_\beta(t) S^y, 
  \]
  where \(M^z_\beta(t)\), \(M^x_\beta(t)\), \(M^y_\beta(t)\) are the 
  unknown parameters to be determined. 
Using  Wick's theorem, as in Refs. \cite{LSM,McCoy1, McCoy_eka}, we have that 
  \begin{equation}
  \label{magnetization_xy}
  M^x_\beta(t) = M^y_\beta(t) = 0.
  \end{equation}  
 Therefore the single-site density matrix of the evolved state is of the form 
  \[
  \rho^1_\beta (t)= \frac{1}{2} I + 2 M^z_\beta(t) S^z,
  \]
 so that 
we are left with determining just the single  parameter 
\[M^z_\beta(t) = \tr\left( S^z \rho^1_\beta(t) \right),\] 
which is the (transverse) magnetization of the system.

 Let us now consider the two-site density matrix \(\rho_\beta^{12}\) of the evolved state. 
  A general    
  two-qubit state is of the form 
 \begin{eqnarray}
 \rho^{12}_\beta (t)&= & \frac{1}{4} I\otimes I + \sum_{j = x, y, z} M^j_\beta(t) (S^j \otimes I 
 + I \otimes S^j ) \nonumber \\
 &+& \sum_{j, k = x, y, z} T^{jk}_\beta(t) S^{j} \otimes S^{k}, 
 \end{eqnarray}
 where \[T^{jk}_\beta (t)= 4 \tr \left(S^j \otimes S^k \rho^{12}_\beta(t)\right)\] are the two-site 
 correlation functions. 
 We already have \(M^x_\beta(t) = M^y_\beta(t) = 0\).
 By applying  Wick's theorem again, one can find that 
 the \(x-z\) and the \(y-z\) correlations are
 vanishing. Therefore, the two-site density matrix of the evolved state is of the form 
 \begin{eqnarray}
 \rho^{12}_\beta (t)& = & \frac{1}{4} I\otimes I +  M^z_\beta(t) (S^z \otimes I 
 + I \otimes S^j ) \nonumber \\
 & + & T^{xy}_\beta(t) (S^x \otimes S^y + S^y \otimes S^x ) \nonumber \\
 & + &\sum_{j = x, y, z} T^{jj}_\beta(t) S^{j} \otimes S^{j}. 
 \end{eqnarray}

To find out the remaining (nonvanishing) parameters of the single and two particle states (\(\rho^1_\beta(t)\) and 
\(\rho^{12}_\beta(t)\) respectively),  explicit use of diagonalizing transformations (Jordan-Wigner,
Fourier, and Bogoliubov transformations) must be made \cite{LSM, McCoy1, McCoy_eka}. 
Using them, one finds that the (transverse)
magnetization of the evolved state is given by \cite{McCoy1}
\begin{eqnarray}
\label{eq_magnetization}
 M^z_{\beta} (t) &=&  \frac{1}{2 \pi} \int_{0}^{\pi} d\phi
\frac{\mbox{tanh}(\frac{1}{2} \beta \Lambda(a))}{\Lambda(a) \Lambda^2 (0)} \nonumber \\
&\times &  \Big\{\left[\cos(2 \Lambda(0) t) \gamma^2 a \sin^2\phi \right] \nonumber \\
&-& \cos \phi \left[(\cos\phi-a)\cos\phi + \gamma^2 \sin^2 \phi \right]\Big\},   \nonumber \\ 
\end{eqnarray}
where \(\Lambda(a)\) and \(\Lambda(0)\) 
are obtained from 
\(
\Lambda(h(t)) = [\gamma^2 \sin^2 \phi + (h(t) - \cos \phi)^2]^{\frac{1}{2}}
\).
The nearest neighbor correlations of the evolved state are given by 
\cite{
McCoy_eka}
\(T^{xy}_\beta (t)= S_\beta(1,t)/i\), \(T^{xx}_\beta (t) = -G_\beta(-1,t)\), \(T^{yy}_\beta (t) = -G_\beta(1,t)\), 
\(T^{zz}_\beta (t)= 4 [M^z_\beta(t)]^2 -  G_\beta(1,t) G_\beta(-1, t) + S_\beta(1,t)S_\beta(-1,t)\), 
where \(G_\beta(R,t)\) and \(S_\beta(R,t)\), for \(R = \pm 1\), are given by 
\begin{eqnarray}
\label{G_R}
 G_\beta(R, t)&=&  \frac{\gamma}{\pi}\int_{0}^{\pi} d \phi \sin(\phi R) \sin\phi 
               \frac{\mbox{tanh}\left(\frac{1}{2} \beta \Lambda(a)\right)}{\Lambda(a)\Lambda^2(0)} \nonumber \\
  \times  \big[\gamma^2 \sin^2 \phi &+& (\cos\phi -a) \cos\phi \nonumber \\
  && \quad \quad \quad \quad  +  a \cos \phi \cos(2\Lambda(0)t) \big] \nonumber \\ 
&-& \frac{1}{\pi} \int_{0}^{\pi} d \phi \cos(\phi R) 
               \frac{\mbox{tanh}\left(\frac{1}{2} \beta \Lambda(a)\right)}{\Lambda(a)\Lambda^2(0)} \nonumber \\
  \times  \big[\{\gamma^2 \sin^2 \phi &+& (\cos\phi -a) \cos\phi \} \cos\phi   \nonumber  \\
  && \quad \quad   -  a \gamma^2 \sin^2 \phi \cos(2\Lambda(0)t) \big], \\ 
S_\beta(R,t) &=&  \frac{\gamma ai}{\pi}\int_{0}^{\pi} d \phi \sin(\phi R) \sin\phi 
               \frac{\sin\left(2t\Lambda(0)\right)}{\Lambda(a)\Lambda(0)}. \nonumber \\
\end{eqnarray}

  \section{Measure of entanglement: Logarithmic negativity}
  \label{entanglement}

  
  Let us now specify the measure of entanglement, which we will use to quantify 
  entanglement of the nearest neighbor spins of our infinite spin chain. 
  There are several ways to  quantify entanglement (see e.g. \cite{MichalQIC, Bennett-assistance}), and in fact there exists no 
``canonical'' entanglement measure. In this paper, we will consider
   logarithmic negativity (LN) \cite{VidalWerner}. It should be stressed, however, that the results do not depend on the choice 
of the entanglement measure. To define 
   logarithmic
  neagativity, let us first introduce negativity.  The negativity \(N(\rho_{AB})\)
  of a bipartite state \(\rho_{AB}\) is defined as the absolute value of the sum of the negative
eigenvalues of  \(\rho_{AB}^{T_{A}}\), where \(\rho_{AB}^{T_{A}}\) denotes the partial
transpose of 
\(\rho_{AB}\) with respect to the \(A\)-part \cite{Peres_Horodecki}. 
The logarithmic negativity
is defined as
  \[
  E_{N}(\rho_{AB}) = \log_2 (2 N(\rho_{AB}) + 1). 
  \]
  In our case, 
the bipartite states are states of two qubits, so that \(\rho_{AB}^{T_{A}}\) has at most 
one negative eigenvalue \cite{Anna-ek}.
Moreover, for two-qubit states, a positive LN implies that the state is 
entangled and distillable \cite{Peres_Horodecki, Horodecki_distillable}, while 
\(E_{N} =0\) implies that the state is separable \cite{Peres_Horodecki}.



\section{Dynamical phase transition  of the \(XY\) spin chain}
\label{sotti}

In this section, we consider the nearest neighbor entanglement of the evolved state 
in the \(XY\) model of the infinite spin chain, at a fixed time \(t\). 
We assume that the 
initial state  is a state of  thermal equilibrium  at zero temperature, in the presence of the transverse field.
We find that the nearest neighbor entanglement
shows a critical behavior, which we call a dynamical phase transition.
Precisely,  the nearest neighbor logarithmic negativity of the evolved state, when 
considered at a given time, shows a criticality as a 
function of the initial transverse field. Moreover, the character of this dynamical phase transition
depends on whether we are near or far from the initial time of disturbance. 
The observed phase transition  is generic, as it occurs
for
a wide range of the anisotropy \(\gamma\).  

Note here that the evolved state at the time \(t\) may be considered as a \emph{stationary state} of the system, 
provided the dynamics is turned off after time \(t\), i.e. the Hamiltonian (\ref{asolH}) is set to zero at time \(t\). 
We stress that turning off the Hamiltonian
at time \(t\) is experimentally feasible \cite{toolbox, proposal_korechhey}.

\begin{figure}[tbp]
\begin{center}
\epsfig{figure= 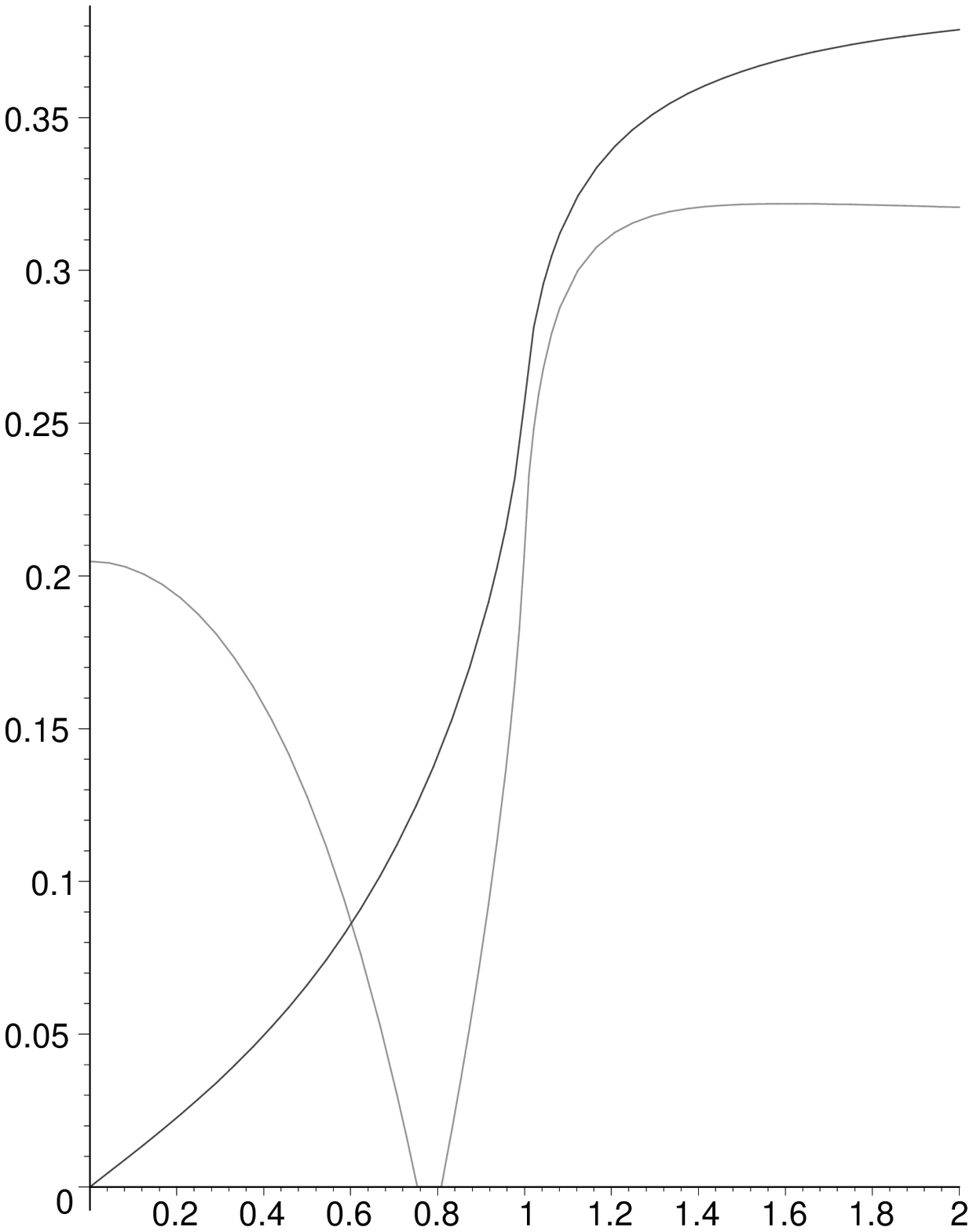,height=0.25 \textheight,width=0.4 \textwidth}
\thicklines
\put(-84,155){\(M^z\)}
\put(-54,128){\(E_N\)}
\put(-164,62){\(E_N\)}
\put(-104,-16){\(a\)}
\caption{
The nearest neighbor logarithmic negativity (\(E_N\)) of the evolved state, 
plotted as a function of the initial transverse field \(a\), at a time that is near the 
time of initial disturbance (\(t=1\)). 
We fix the
anistropy at \(\gamma = 0.5\).  The initial state of evolution is a state of thermal equilibrium at 
zero temperature. \(E_N\) vanishes at a critical value \(a_c\) and revives at \(\overline{a}_c\).
We will show in Sec. \ref{sec_asol} that for values of the initial 
field in the region near the phase 
transitions, entanglement behaves nonmonotonically with respect to the temperature of the initial  equilibrium state. 
The transverse magnetization (\(M^z\)) of the evolved state is also plotted; it does not 
show a similar critical behavior as a function of \(a\).
}
\label{ent_wrta_time1}
 \end{center}
\end{figure}

Let us first consider the behavior of the nearest neighbor entanglement with respect to 
the initial transverse field \(a\), at a time \(t\) that is near the initial time of disturbance. 
In Fig. \ref{ent_wrta_time1}, we plot the nearest neighbor LN of the evolved state with respect 
to \(a\), at \(t=1\), and for the anisotropy \(\gamma = 0.5\). 
Entanglement exhibits  criticalities, as the system parameter \(a\), i.e. the initial transverse 
field is  changed: \(E_N\) vanishes at a critical value \(a_c\) and revives at another critical value \(\overline{a}_c\).
 Note that a similar phenomenon is absent for magnetization. We will see in the
succeeding section that  for values of the initial field that is close to the critical regions, 
entanglement of the evolved state behaves nonmonotonically as a function of the 
temperature of the initial equilibrium state.

Similar DPT's of entanglement, as the system parameter \(a\) changes,
can be seen for other values of the time \(t\), sufficiently near 
to the initial moment of disturbance, as well as for other values of the anisotropy \(\gamma\).

\begin{figure}[tbp]
\begin{center}
\epsfig{figure= 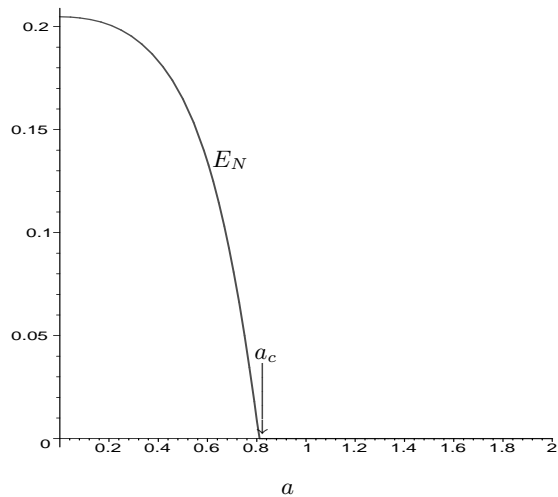,height=0.25 \textheight,width=0.4 \textwidth}
\put(-130,108){\(E_N\)}
\put(-104,-16){\(a\)}
\put(-114, 35){\(a_c\)}
\put(-114, 18){\(\Biggl\downarrow\)}

\caption{
The nearest neighbor logarithmic negativity (\(E_N\)) of the evolved state \(\rho^{12}_\beta(t)\) is plotted
against the initial transverse field \(a\), at \(t =10\), which is comparatively large, as 
compared to that of Fig. \ref{ent_wrta_time1}.  
We again fix the
anistropy  \(\gamma = 0.5\), and \(\beta \rightarrow \infty\), as in Fig. \ref{ent_wrta_time1}. 
Here
we observe again a kind of criticality, but 
of a significantly different character than the one 
in Fig. \ref{ent_wrta_time1}. 
In  Section  \ref{sec_asol}, 
 we show that 
entanglement behaves monotonically as a function of temperature of the initial 
equilibrium state, for \(a \approx a_c\) in this regime of \(t\)'s.}
\label{ent_wrta_t2}
 \end{center}
\end{figure}

However, as the time grows,
the nature 
of the DPT changes significantly. 
In Fig. \ref{ent_wrta_t2}, we plot the nearest neighbor LN for a time that is comparatively far away from 
\(t=0\),
against the 
initial field \(a\). 
Again, a 
dynamical phase transition is observed, but  one 
that is 
 different from the one observed in Fig. \ref{ent_wrta_time1}.
In Fig. \ref{ent_wrta_t2}, the system undergoes a phase transition from the entangled phase
to the separable phase at 
\(a_c\), but no re-entrance behavior is observed.
Moreover, in the suceeding section, we show that the nonmonotonicity of entanglement
is no longer present in this case of large \(t\). 
A signature of dynamical phase transition is absent for magnetization for both small and large \(t\).

\begin{figure}[tbp]
\begin{center}
\epsfig{figure= 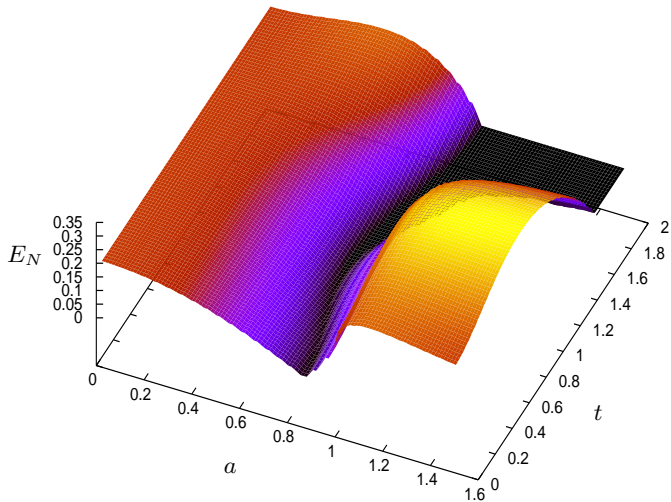,height=0.35 \textheight,width=0.58 \textwidth}
\put(-204,26){\(a\)}
\put(-64,46){\(t\)}
\put(-286,106){\(E_N\)}
\caption{
The nearest neighbor logarithmic negativity (\(E_N\)) of the evolved state \(\rho^{12}_{\beta \rightarrow \infty}(t)\) is plotted
against the initial transverse field \(a\) and time \(t\) for the anisotropy \(\gamma = 0.5\).
On the \(E_N = 0\) plane, there are two curves of DPTs, both of which start at around \(\{a = 0.8, t =0\}\), and then they diverge off
with increasing \(t\), forming a ``river'' of separable states between themselves. The \(t=1\) and \(t =10\) slices of this surface were already 
discussed before. For the \(t=0\) slice, entanglement is seen to vanish, as it should, as \(a\) grows. Note that this is different from the 
\(t=1\) slice behavior, where the entanglement converges to a positive value.}
\label{fig_konark}
 \end{center}
\end{figure}

\begin{figure}[tbp]
\begin{center}
\epsfig{figure= 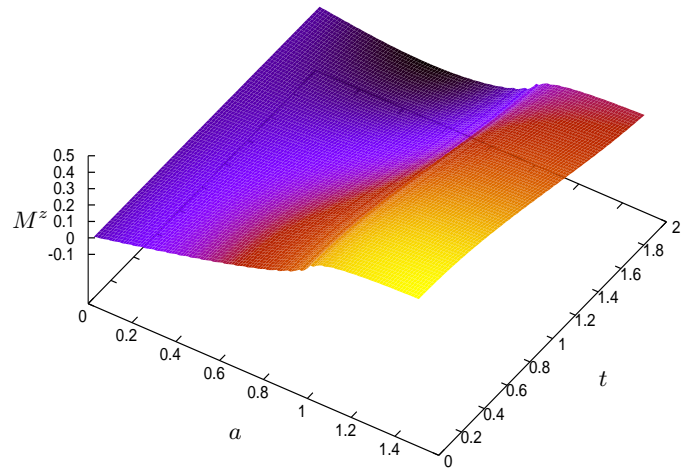,height=0.35 \textheight,width=0.58 \textwidth}
\put(-204,26){\(a\)}
\put(-64,46){\(t\)}
\put(-286,106){\(M^z\)}
\caption{
The magnetization (\(M^z\)) of the evolved state \(\rho^{12}_{\beta \rightarrow \infty}(t)\) is plotted
against the initial transverse field \(a\) and time \(t\) for the anisotropy \(\gamma = 0.5\).
}
\label{fig_khajuraho}
 \end{center}
\end{figure}

To obtain a global perspective of the behavior of entanglement,  we plot it 
with respect to both \(t\) and \(a\), at a fixed value of the anisotropy \(\gamma\) (\(\gamma = 0.5\))
in Fig. \ref{fig_konark}.
Note that a similar behavior is absent in magnetization, as seen in Fig. \ref{fig_khajuraho}.

An interpretation of the results for small and large \(a\) (as compared to the region of 
phase transitions) is the following: For small \(a\), the initial state is entangled, 
and its entanglement survives even for long times \(t\) (see Fig. \ref{ent_wrta_t2}). 
On the other hand, for large \(a\), the
initial state is close to a separable state. This state is however, very different from any eigenstate 
of the Hamiltonian, and is thus very strongly affected by the dynamics, generating significant entanglement for small times \(t\).
The case of large \(a\) is therefore similar to that in Refs. \cite{Briegel_orey_orey, Briegel_ebong}, where 
the time-independent Ising Hamiltonian with interaction in the \(z\)-direction, is made to act on a product of eigenstates of 
\(\sigma_x\).

Note here that in contrast to the case of small \(a\),   
there is no revival of entanglement with respect to time \(t\), for large \(a\) and large \(t\) (compare  Fig. \ref{ent_wrta_time1} and 
Fig. \ref{ent_wrta_t2}). 
This feature is different from that in the Ising Hamiltonian, or from that in 
spin glass \cite{rotating-vase}, where the state re-enters to the entangled and separable phases again and again after certain time intervals.  
The continuous character of the spectrum of the infinite \(XY\) chain is most probably responsible for this effect: The
dynamics is mixing the states in such a way that for large \(a\), the revival of entanglement is possible only for relatively short 
  times \(t\), or alternatively, that for large \(t\), there is no entanglement at large \(a\).

  \section{Nonmonotonicity of entanglement with temperature}
  \label{sec_asol}

  In the preceding section, we have obtained the 
DPT
of nearest neighbor entanglement of 
  the evolved state of the infinite spin chain.
The DPT was controlled by 
the transverse field \(a\). 
The initial state of the evolution, however,  was taken to be the equilibrium state at zero temperature. 
  In this section, we will study this 
criticality
and  the monotonicity of
  nearest neighbor entanglement, considered as a function
  of the temperature of the initial equilibrium state.

  For definiteness, consider the dynamical phase transition 
of 
Fig. \ref{ent_wrta_time1},
observed for the nearest neighbor entanglement of the evolved state at time \(t = 1\), and 
  for the anisotropy \(\gamma= 0.5\). The evolution 
there had started from the equilibrium state at zero temperature. Consider now the evolution 
  in which  
  the initial state 
  is the equilibrium state \(\rho^{eq}_\beta\), at a certain temperature \(T = 1/\beta\). 
  We again look at the nearest neighbor entanglement of the evolved state at time \(t=1\) and 
  for anisotropy \(\gamma = 0.5\), but now 
  as a function of the temperature \(T\) of the initial equilibrium state, and for a given value of 
  the initial transverse field. It turns out that the behavior of entanglement (with respect to temperature) is 
   qualitatively different, depending on whether we are near or far away from the dynamical 
  phase transition of entanglement in Fig.   \ref{ent_wrta_time1}. 
 We find that it is possible to obtain three qualitatively different regions
  of the transverse field \(a\), according to the behavior of entanglement with respect to the 
  temperature of the initial equilibrium state. 

\begin{enumerate}

\item[(i)] \emph{The initial transverse field \(a\) is far away (either lower or higher) 
           from the critical regions}: In this case, the nearest neighbor LN is 
	   monotonic with respect to the temperature of the initial equilibrium state.  
	   As temperature is lowered, LN increases monotonically, and ultimately converges to a nonzero value.
	   This is illustrated in Fig. \ref{ent_beta_apoint5}, for an exemplary value of \(a = 0.5\), 
	   that is comparatively far away from the critical region (in comparison to the cases 
	   considered in item (iii)).
\begin{figure}[tbp]
\begin{center}
\epsfig{figure= 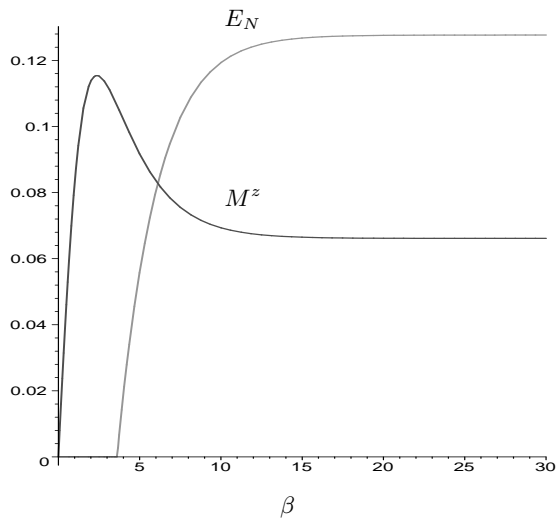,height=0.25 \textheight,width=0.4 \textwidth}
\put(-125,168){\(E_N\)}
\put(-125,100){\(M^z\)}
\put(-104,-16){\(\beta\)}
\caption{ 
The logarithmic negativity \(E_N\) of the state \(\rho^{12}_\beta(t)\),  is plotted as a function of 
the inverse temperature \(\beta\) of the initial equilibrium state \(\rho^{eq}_\beta\). We choose \(t=1\)
and \(\gamma = 0.5\), just as in Fig. \ref{ent_wrta_time1}. The transverse field \(a\) is chosen to be \(0.5\), 
which is supposed to be comparatively far away from the critical regions in Fig. \ref{ent_wrta_time1} (in comparison
to the values of \(a\) used in Figs. \ref{ent_beta_apoint74} and \ref{ent_beta_apoint8} below). We find that entanglement increases
monotonically with decreasing \(T\) (increasing with \(\beta\)). For reference and comparison, we also plot the transverse magnetization 
\(M^z\)
of \(\rho_\beta(t)\). 
}
\label{ent_beta_apoint5}
 \end{center}
\end{figure}

\item[(ii)] \emph{The initial field is within the separable phase}:  The nearest neighbor LN is
nonmonotonic with respect to the temperature of the initial equilibrium state. In particular, 
there are regions of temperature for which entanglement increases with increasing temperature
(see also Ref. \cite{Scheel} in this respect).  
A plot of nearest neighbor LN with respect to the initial temperature, 
is given in Fig. \ref{ent_beta_apoint78}, for an exemplary value of 
the transverse field in the separable phase. 
\begin{figure}[tbp]
\begin{center}
\epsfig{figure= 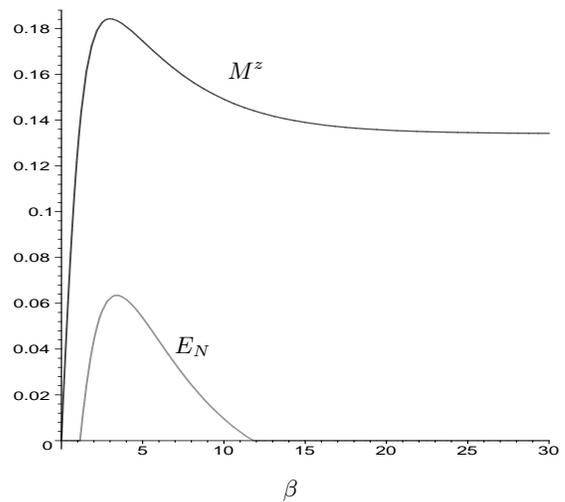,height=0.25 \textheight,width=0.4 \textwidth}
\put(-145,38){\(E_N\)}
\put(-125,142){\(M^z\)}
\put(-104,-16){\(\beta\)}
\caption{
The logarithmic negativity \(E_N\) of the state \(\rho^{12}_\beta(t)\),  is plotted as a function of 
the inverse temperature \(\beta\) of the initial equilibrium state \(\rho^{eq}_\beta\). We choose \(t=1\)
and \(\gamma = 0.5\), just as in Fig. \ref{ent_wrta_time1}. The transverse field \(a\) is chosen to be \(0.78\), 
which is within the separable phase in Fig. \ref{ent_wrta_time1}. We find that entanglement is nonmonotonic
with respect to temperature. In particular, therefore, there is range of temperature, for which
entanglement is increasing with increasing temperature. 
For sufficiently low \(T\) (and, as expected, for high \(T\)), 
entanglement 
is vanishing, in contrast to the situation in Figs. \ref{ent_beta_apoint74} and \ref{ent_beta_apoint8} below. 
Again, for reference and comparison, we also plot the transverse magnetization 
\(M^z\)
of \(\rho_\beta(t)\). 
}
\label{ent_beta_apoint78}
 \end{center}
\end{figure}
Note that entanglement in this case is nonvanishing only for 
moderate values of \(T\). 
For very high and very low \(T\), entanglement vanishes. This is different than in item (iii) below. 

\item[(iii)] \emph{The initial field is in the critical region of the entangled phase}: If the transverse field
is sufficiently close to the critical region, but still being in the entangled phase,  the nearest neighbor LN is 
again nonmonotonic with respect to the temperature of the initial state. 
However, the added feature is that the entanglement 
converges to a nonvanishing value for low \(T\). 
In case (ii), the entanglement is vanishing for sufficiently low \(T\) (and 
hence for sufficiently large \(\beta\)) (see Fig. \ref{ent_beta_apoint78}). For sufficiently high \(T\),
entanglement is of course again vanishing, just as in the items (i) and (ii) above.
 The nearest neighbor LN, plotted for two exemplary values 
of  \(a\) in the region under consideration, are given in Figs. \ref {ent_beta_apoint74}
and \ref{ent_beta_apoint8}.
\begin{figure}[tbp]
\begin{center}
\epsfig{figure= 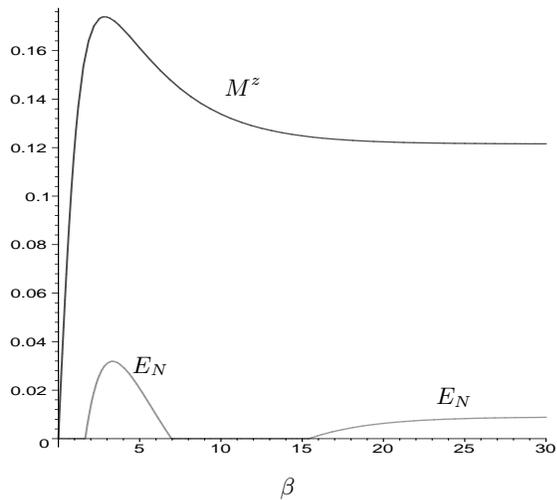,height=0.25 \textheight,width=0.4 \textwidth}
\put(-125,135){\(M^z\)}
\put(-45,18){\(E_N\)}
\put(-160,30){\(E_N\)}
\put(-104,-16){\(\beta\)}
\caption{The logarithmic negativity \(E_N\) of the state \(\rho^{12}_\beta(t)\),  is plotted as a function of 
the inverse temperature \(\beta\) of the initial equilibrium state \(\rho^{eq}_\beta\). We choose \(t=1\)
and \(\gamma = 0.5\), just as in Fig. \ref{ent_wrta_time1}. The transverse field \(a\) is chosen to be \(0.74\), 
which is in a critical region in Fig. \ref{ent_wrta_time1}, but in the entangled phase (the first entangled 
phase). 
Just as in Fig. \ref{ent_beta_apoint78},  we find that entanglement is nonmonotonic
with respect to temperature. However, in contrast to the case in Fig. \ref{ent_beta_apoint78}, 
entanglement is nonvanishing for low \(T\). For sufficiently low \(T\), entanglement converges to a 
nonvanishing value.
For comparison, we also plot the transverse magnetization 
\(M^z\)
of \(\rho_\beta(t)\).
}
\label{ent_beta_apoint74}
 \end{center}
\end{figure}
\begin{figure}[tbp]
\begin{center}
\epsfig{figure= 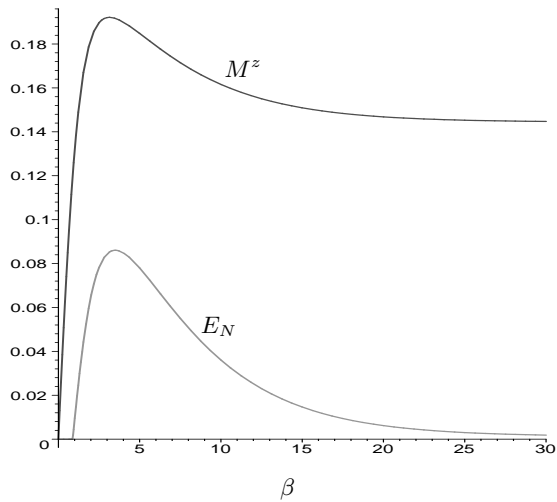,height=0.25 \textheight,width=0.4 \textwidth}
\put(-125,142){\(M^z\)}
\put(-134,45){\(E_N\)}
\put(-104,-16){\(\beta\)}
\caption{
The logarithmic negativity \(E_N\) and the transverse magnetization \(M_z\) 
of the state \(\rho^{12}_\beta(t)\),  are plotted as  functions of 
the inverse temperature \(\beta\) of the initial equilibrium state \(\rho^{eq}_\beta\). We choose \(t=1\)
and \(\gamma = 0.5\), just as in Fig. \ref{ent_wrta_time1}. The transverse field \(a\) is chosen to be \(0.81\), 
which is again (just as in Fig. \ref{ent_beta_apoint74}) 
in a critical region of Fig. \ref{ent_wrta_time1}, but in the entangled phase (the \emph{second} entangled 
phase). 
The qualitative features of \(E_N\) are just as in Fig. \ref{ent_beta_apoint74} above. 
}
\label{ent_beta_apoint8}
 \end{center}
\end{figure}

\end{enumerate}

We stress that although we have considered here the case 
only for
\(\gamma= 0.5\), the results are generic and have been numerically checked 
 for several values of \(\gamma\).

It is to be noted that the behavior of transverse magnetization does not seem to alter considerably
as we pass from one entangled phase to another, through the separable phase, as is seen in 
Figs. \ref{ent_beta_apoint5}, \ref{ent_beta_apoint78}, \ref{ent_beta_apoint74}, \ref{ent_beta_apoint8}.   
The absence of this  feature in magnetization, and its presence in entanglement, once again underlines 
that complexity of physical phenomena can be understood in terms of entanglement.

With respect to the nonmonotonicity of entanglement with the temperature of the initial state, let us note here 
that the usual intuition is that entanglement is a fragile quantity, and therefore it decays with noise. 
It is also usual to see an increase of temperature as a model of increase of noise in the system. 
This is for instance corroborated here in Figs. 
\ref{ent_beta_apoint5}, \ref{ent_beta_apoint78}, \ref{ent_beta_apoint74}, \ref{ent_beta_apoint8},
where entanglement vanishes for sufficiently large temperatures. 
However, we see here that for moderate values of \(T\), 
the fragility of entanglement is a more complex issue. There can be ranges of temperatures for which
entanglement actually increases with temperature.

Until now, we have been discussing the temperature effects for 
the dynamical phase transitions that were exemplified in Fig. \ref{ent_wrta_time1}
of the preceding section. A different sort of 
DPT
was also obtained in the preceding section, as 
exemplified in Fig. \ref{ent_wrta_t2}. Surprisingly, in this case, the nearest neighbor entanglement 
does not behave 
as in the case of Fig. \ref{ent_wrta_time1}.

\begin{figure}[tbp]
\begin{center}
\epsfig{figure= 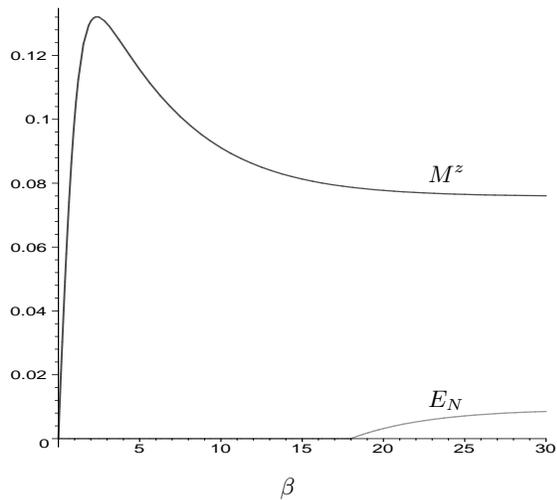,height=0.25 \textheight,width=0.4 \textwidth}
\put(-48,102){\(M^z\)}
\put(-48,17){\(E_N\)}
\put(-104,-16){\(\beta\)}
\caption{The logarithmic negativity \(E_N\) of the state \(\rho^{12}_\beta(t)\),  is plotted as a function of 
the inverse temperature \(\beta\) of the initial equilibrium state \(\rho^{eq}_\beta\). We choose \(t=10\)
and \(\gamma = 0.5\), just as in Fig. \ref{ent_wrta_t2}. The transverse field \(a\) is chosen to be \(0.8\), 
which is approximately just the point of phase transition in Fig. \ref{ent_wrta_t2}. 
Although we are almost on the point of phase transition, we find that in contrast to the case of near-time 
phase trasnition,  entanglement is monotonic
with respect to temperature.
For comparison, we also plot the transverse magnetization 
\(M^z\)
of \(\rho_\beta(t)\).
}
\label{ent_beta_apoint8_t10}
 \end{center}
\end{figure}

In Fig. \ref{ent_wrta_t2}, we plotted the nearest neighbor LN of the evolved state at time \(t =10\), 
which is comparatively far away from the point of initial disturbance in the transverse field. 
The plot was with respect to the system parameter \(a\), 
and a criticality was obtained at \(a \approx 0.8\), for  the anisotropy \(\gamma = 0.5\).
We 
consider now the nearest neighbor entanglement of the evolved state at the time \(t =10\) and for \(\gamma = 0.5\),
as a function of the temperature of the initial equilibrium state.   
As we see, in contrast to the case of the phase transitions in Fig. \ref{ent_wrta_time1}, 
the nearest neighbor entanglement does not change its behavior as we choose different values of the 
transverse field \(a\). 
In particular, the nearest neighbor LN of the evolved state is monotonic with temperature, and converges 
to a nonvanishing value for low \(T\) (large \(\beta\)). In Fig. \ref{ent_beta_apoint8_t10}, we plot the nearest neighbor 
LN with respect to the initial temperature, for \(a = 0.8\),  \(t = 10\),  \(\gamma = 0.5\). 
However this 
feature is generic. We have also obtained similar features for several values of \(\gamma\).

In Ref. \cite{ent_length}, the authors define an entanglement length (range of quantum correlations), 
which is shown to diverge at the critical points for a wide range of spin systems. The definition is in terms 
of a quantity called localizable entanglement, which is usually hard to compute. However, there is a useful upper bound of this quantity 
in terms of the entanglement of assistance \cite{Bennett-assistance}. Considering this upper bound for the case of 
the nearest neighbor density matrix, as well as for the next-nearest neighbor density matrix, we have checked that  such definition of entanglement 
length does not seem to be able to characterise the dynamical phase transitions discussed in this paper. This indicates that the quantum phase 
transitions considered in Ref. \cite{ent_length} are of a different character from the ones discussed here. 
Moreover, the behavior of entanglement with temperature of the system, can be seen as an independent candidate for understanding the 
phase transitions in the system.

  \section{Discussion}
  \label{discussion}
  
In this paper, we have investigated the dynamics of entanglement in the evolution of the 
infinite asymmetric \(XY\) spin chain,
in
an initial transverse field.
 One motivation behind our study is that the dynamics of
entanglement in the evolution of many-body spin-systems have been used 
to implement quantum computation and short range quantum communication \cite{Briegel_orey_orey, Sougato}. 
We also hope to be able to understand the physical phenomena in complex systems with the 
help of entanglement \cite{Osterloh, Nielsen, ent_length, numerical}.

For short times, we found a critical behavior of nearest neighbor 
entanglement of the system, with respect to the initial
transverse field. The nearest neighbor entanglement 
vanishes for a certain value of the initial transverse field, 
to enter into the separable phase from an entangled phase. 
For a higher value of the field, there is a revival of entanglement, and the system re-enters 
into the entangled phase.
For long times, there is again a criticality as the system moves from an entangled phase to a separable 
phase. However, there is no re-entrance into the entangled phase.  
In both cases studied, the system evolved from an initial 
(thermal) equilibrium state at zero temperature, and then we consider the nearest neighbor entanglement of the system 
at a fixed time. 
We refer to the regions of the transverse field, where the transition from entangled phases to separable phases occur,
as the critical regions.


Surprisingly, we have shown that the
nearest neighbor entanglement is nonmonotonic with respect to temperature in these critical regions,
for short times. 
Similar behavior can also be
seen in the separable phase. However for values of the transverse field that is deep inside the entangled phases, 
entanglement 
is strictly decreasing
with temperature,  both for short and long times.


Finally, let us note that it is 
important to consider the behavior of entanglement with respect to temperature in many body systems, as 
one of the main challenges in implementing quantum information processing tasks is to preserve entanglement 
in a noisy environment. 
Temperature is a usual intuitive way to model noise in such systems. Our findings indicate 
that the behavior of entanglement with respect to temperature, at least for moderate 
values of temperature, is quite complex. In particular, we found that for some ranges of temperature, 
entanglement in the system can grow with increasing temperature.
It is interesting to look for similar nonmonotonic behavior of entanglement with respect 
to noise in the system,
in other physical models, to find out  how general such behavior can be. 

\acknowledgments

We thank Jens Eisert for discussions at the 5th European QIPC   Workshop 2004, in Rome.
  We acknowledge support of the Deutsche Forschungsgemeinschaft 
(SFB 407, SPP 1078, SPP 1116, 432POL), the 
Alexander von Humboldt  
Foundation, 
the EC Contract No. IST-2002-38877 QUPRODIS, the ESF Program QUDEDIS, and EU IP SCALA.

\end{document}